\documentclass[reprint,amsmath,amssymb,aps,pra]{revtex4-2}

\usepackage[english]{babel}
\usepackage{amsfonts}
\usepackage{bm}
\usepackage{bbm}
\usepackage{physics}

\usepackage{graphicx}
\graphicspath{ {./Figures/} }
\usepackage[small,tight,FIGTOPCAP]{subfigure}

\usepackage{xcolor}
\usepackage{hyperref}
\hypersetup{colorlinks=true, citecolor=blue, linkcolor=black, urlcolor=blue}

\begin{document}
\title{\texorpdfstring{$\mathcal{PT}$}{PT}-Symmetry breaking in quantum spin chains with exceptional non-Hermiticities}
\author{Jacob Muldoon and Yogesh N. Joglekar}
\affiliation{Department of Physics, Indiana University-Purdue University Indianapolis (IUPUI), Indianapolis, Indiana 46202}

\begin{abstract}
    Since the realization of quantum systems described by non-Hermitian Hamiltonians with parity-time ($\mathcal{PT}$) symmetry, interest in non-Hermitian, quantum many-body models has steadily grown. 
    Most studies to-date map to traditional quantum spin models with a non-Hermiticity that arises from making the model parameters complex or purely imaginary. 
    Here, we present a new set of models with non-Hermiticity generated by splitting a Hermitian term into two Jordan-normal form parts , and the perturbations are confined to one or two sites. 
    We present exact diagonalization results for the $\mathcal{PT}$-threshold in such models, and provide an analytical approach for understanding the numerical results. 
    Surprisingly, with non-Hermitian potentials confined to two or even a single site, we find a robust $\mathcal{PT}$ threshold that seems insensitive to the size of the quantum spin chain. 
    Our results provide a pathway to experimentally feasible non-Hermitian quantum spin chains where the confluence of many-body effects and non-Hermiticity effects can be observed.
\end{abstract}

\maketitle


\section{Introduction}
\label{sec:intro}

Since the seminal discovery of Bender and co-workers 25 years ago~\cite{Bender1998}, the field of non-Hermitian systems has dramatically flourished. 
Research initially focused on continuum, non-relativistic Schr\"{o}dinger equations with complex (often, purely imaginary) potentials that were invariant under combined operations of parity and time-reversal, i.e. $V(x)=V^*(-x)$~\cite{Bender2002,Mostafazadeh2002,Mostafazadeh2003}. 
Such $\mathcal{PT}$-symmetric Hamiltonians showed purely real spectra at small non-Hermiticity, going over to complex-conjugate spectra at large non-Hermiticity~\cite{Bender2007,Joglekar2013}. 
Experiments in wave systems (optics~\cite{Guo2009,Rueter2010,Regensburger2012}, acoustics~\cite{Zhu2014}, and the like~\cite{Peng2014,Hodaei2014}) with balanced, spatially separated gain and loss, provided a simple physical interpretation for $\mathcal{PT}$-symmetric Hamiltonians as effective models for open systems~\cite{Schindler2011,Bender2013}. 
From this vantage point, the $\mathcal{PT}$-symmetry breaking transition marks the concomitant emergence of amplifying and decaying modes in an open system. 
Thus, in the classical domain, $\mathcal{PT}$-symmetric Hamiltonians are often modeled with purely anti-Hermitian potentials that signify local amplification or absorption. 
Over the years, these ideas have been generalized to time-periodic models~\cite{Joglekar2014,Chitsazi2017,QuirozJuarez2022}, non-Markovian models~\cite{Cochran2021,Wilkey2020}, and synthetic degrees of freedom~\cite{Zhang2020,Ding2021a}, all in the classical domain.

In the quantum domain, creation of balanced gain and loss potentials is precluded by thermal fluctuations associated with the dissipation~\cite{Kubo1966}, and even at zero temperature, the quantum noise associated with linear amplifiers~\cite{Caves1982,Scheel2018}. 
Instead, the coherent, non-unitary dynamics generated by $\mathcal{PT}$-symmetric Hamiltonians is simulated by mode-selective losses~\cite{Xiao2017,Li2019}, Hamiltonian dilation~\cite{Wu2019}, or unitary dilation~\cite{Maraviglia2022} methods. 
Most recently, it was realized that a Lindbladian, minimal quantum system~\cite{Lindblad1976,Gorini1976,Manzano2020}, when post-selected on trajectories that do not undergo quantum jumps~\cite{Moelmer1993,Minganti2019}, is described by a non-Hermitian, $\mathcal{PT}$-symmetric Hamiltonian with state-dependent, trace-preserving non-linearity~\cite{Brody2012}. 
This technique has enabled the exploration of non-Hermitian Hamiltonians in quantum, two-level systems~\cite{Naghiloo2019,Klauck2019,Ding2021,Quinn2023}. 

Theoretical studies of non-Hermitian, quantum many-body models have commenced by changing parameters in their Hermitian counterparts from real to complex while maintaining their functional form~\cite{Korff2008,CastroAlvaredo2009,Deguchi2009,Ashida2017,Yamamoto2019,ZhangJin2020,Liu2021,Lenke2021a,Bi2021,Yamamoto2022,Chen2023}. Such models inherit the symmetries of their Hermitian counterparts, such as translational invariance, and therefore can be analytically investigated. However, preserving those symmetries comes at the cost of having non-Hermitian potentials on a large number of sites, spins, or other relevant degrees of freedom. Experimentally, observing non-Hermitian dynamics in even a single qubit is constrained by an exponentially decaying probability $p_1\sim e^{-\gamma t}$ for obtaining no-quantum-jump trajectories~\cite{Naghiloo2019,Klauck2019,Ding2021,Quinn2023}. When simulating the dynamics of a system with $n$ non-Hermitian qubits, the success probability $P_n$, {\it given by quantum trajectories where none of them undergoes a quantum jump}, is doubly-exponentially suppressed, $P_n=p_1^n\sim\exp(-n\gamma t)$. This experimental-feasibility analysis endorses a minimal footprint for the non-Hermitian potential, even at the expense of symmetries.

Here we present a class of models with non-Hermiticity created by splitting a Hermitian potential into two, Jordan-form terms and then spatially separating them. For example, in a transverse field quantum Ising model, this means $\gamma\sigma^x_m\rightarrow \gamma(\sigma^+_{m-n}+\sigma^{-}_{m+n}$) where $\sigma^\alpha_m$ represents the relevant Pauli operator on site $m$. Note that $\sigma^{\pm}\equiv (\sigma^x\pm i\sigma^y)/2$ are rank-1, Jordan-form matrices i.e. they represent single-qubit Hamiltonians at an exceptional point (EP)~\cite{Naghiloo2019}. Contrary to typical anti-Hermitian potentials, (exceptional) non-Hermiticities such as $\gamma\sigma^\pm$, with their EP degeneracies do not have an energy scale. In quantum spin systems with finite number of levels, the mapping between raising/lowering operators $\sigma^\pm$ and gain/loss is ambiguous due to the presence of a ceiling in the spectrum. 
On the contrary, in bosonic models such as two coupled oscillators, this splitting procedure will generate non-Hermitian gain/loss potentials such as $\gamma(a^\dagger_1+a_2)$. We emphasize that the operators $\sigma^{\pm}$ are terms in the Hamiltonian, not dissipators routinely used in Lindblad dynamics to model spontaneous emission and absorption; the latter give rise to anti-Hermitian potentials~\cite{Avila2020}. 

In this paper, we investigate the $\mathcal{PT}$-symmetry breaking threshold in transverse field Ising models with finite number of spins $N$ and its dependence on parameters using exact diagonalization method. 
Other traditional techniques such as perturbation theory or tensor networks are ideal for probing a small, ground-state-proximate subspace of the exponentially-large Hilbert space. 
Determining the $\mathcal{PT}$-breaking threshold---where the Hamiltonian first develops complex-conjugate eigenvalues --- requires knowledge of the entire spectrum, since the states that develop complex eigenvalues are typically not at the bottom (or the top) of the band~\cite{Liang2014}. 

The plan of the paper is as follows. 
In Sec.~\ref{sec:model} we introduce the canonical quantum Ising chain and its non-Hermitian variations. 
The non-Hermitian variations on it consist of perturbations on one or two sites. 
In addition to the $\mathcal{PT}$-threshold, we also present the flow of eigenvalues across the $\mathcal{PT}$-symmetry breaking transition. 
In Sec.~\ref{sec:analytical} we present a simple analytical approach that explains the surprisingly robust $\mathcal{PT}$ threshold results from Sec.~\ref{sec:model}. 
We conclude the paper in Sec.~\ref{sec:disc} with higher-spin generalizations, brief feasibility analysis, and summary. 
The $\mathcal{PT}$-threshold results are valid for chains with $N>2$ where the bulk-vs-edge sites and periodic-vs-open boundary conditions are unambiguously defined, but do not depend on $N$. 


\begin{figure*}
    \centering
    \includegraphics[width=\textwidth]{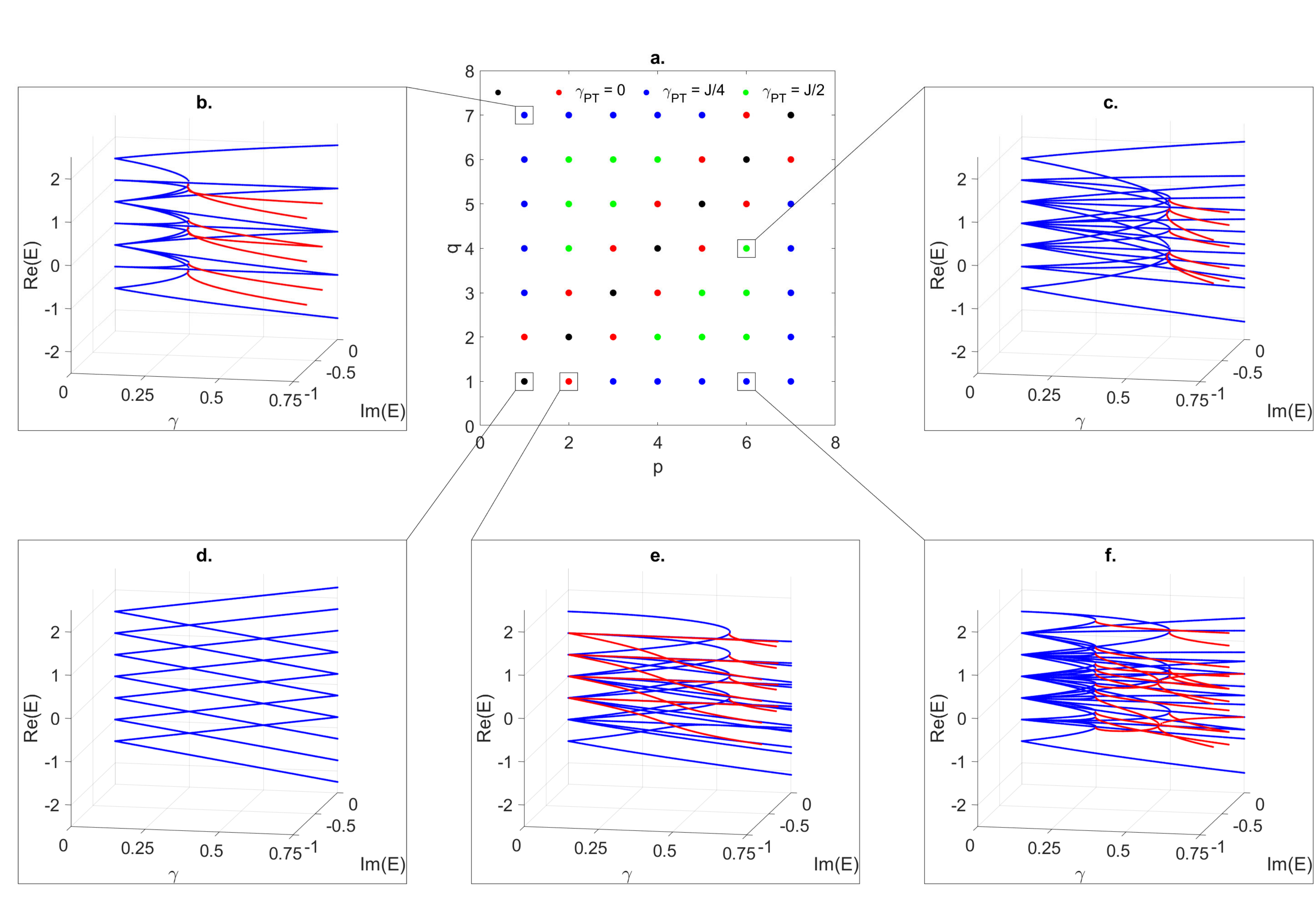}
    \caption{
        $\mathcal{PT}$-breaking threshold for a 7-spins chain with $h_z=0$ and non-Hermiticity $\Gamma^{+}_{p,q}$, Eq.(\ref{eq:G}). 
        (a) Apart from the Hermitian case at $p=q$ (black circles), threshold takes three possible values: zero for adjacent sites (red circles); $\gamma_{\textrm{PT}}(p,q)=J/4$ when at least one site is at the edge (blue circles); and $\gamma_{\textrm{PT}}(p,q)=J/2$ when both sites are in the bulk. 
        (b)-(f) show the flow of eigenvalues $\Re(E)(\gamma)/J$ (blue lines) and $\Im(E)(\gamma)/J>0$ (red lines) for $p,q$ locations marked in (a). 
        At $\gamma=0$, the system has 7 particle-hole symmetric bands with varying degeneracies spanning the $2^7=128$ eigenvalues; at a finite $\gamma$, the particle-hole symmetry is generally broken. 
        Ground-state band typically does not participate in the $\mathcal{PT}$-symmetry breaking transition. 
        Therefore, variational or perturbative methods that focus on the lowest-lying states cannot be used to determine the $\mathcal{PT}$-symmetry breaking threshold.
    }
    \label{fig:allpanels}
\end{figure*}

\section{Non-Hermitian quantum Ising models}
\label{sec:model}

The canonical quantum Ising model with $N$ sites is described by the Hamiltonian 
\begin{align}
    \label{eq:h0}
    H_0(J,h_z) &= -\frac{J}{4}\sum_{i=1}\sigma^{x}_{i}\sigma^{x}_{i+1} - \frac{h_z}{2}\sum_{i=1}\sigma^{z}_{i}
\end{align}
where $J>0$ is the ferromagnetic coupling between adjacent spins, the uniform transverse field is along the $z$-axis, and the boundary term $\sigma^x_N\sigma^x_1$ is included when periodic-boundary conditions are required~\cite{Lieb1961,Pfeuty1970}. 
This exactly solvable model undergoes a quantum phase transition from a spontaneously-broken $\mathbb{Z}_2$-symmetry phase to a paramagnetic phase with short-range correlations as the transverse field strength crosses $h_z=J/2$~\cite{Sachdev2009,Fradkin2013}. 

In this section, we investigate its varied non-Hermitian extensions. We start with the $h_z=0$ case where $H_0$ contains only commuting operators and can be trivially diagonalized. Adding a non-commuting term to $H_0$ changes it into a genuine quantum Ising model.


\subsection{Two-site Perturbations with Hermitian or anti-Hermitian limit \texorpdfstring{($h_z=0$)}{}}
\label{subsec:two}

Consider the non-Hermitian extension
\begin{align}
    \label{eq:heff}
    H_\textrm{eff}(J,h_z|\gamma)&= H_0(J,h_z)+\Gamma^{+}_{pq}(\gamma),\\
    \label{eq:G}
    \Gamma^{+}_{pq}&=\gamma(\sigma_{p}^{+}+\sigma^{-}_{q})\neq \Gamma^{+\dagger}_{pq},
\end{align}
where $\gamma>0$ is the strength of the exceptional perturbations $\sigma^\pm$, and $1\leq p,q\leq N$ denote their locations along the chain.
When $p=q$, the perturbation Eq.(\ref{eq:G}) is trivially Hermitian and the system has no threshold. 
Since $H_\textrm{eff}$ has real entries, its characteristic polynomial has real coefficients and its eigenvalues are real or complex conjugates~\cite{Ruzicka2021}. 

Figure~\ref{fig:allpanels} summarizes the $\mathcal{PT}$-threshold phase diagram of such quantum spin chain in the absence of transverse field. 
It involves calculating the spectrum of $H_{\textrm{eff}}(J,0|\gamma)$ by exact diagonalization, and then recursively increasing the strength of $\Gamma^{+}$ until complex-conjugate eigenvalues emerge at the threshold $\gamma_{\textrm{PT}}$. 
Figure~\ref{fig:allpanels}(a) shows the dimensionless threshold $\gamma_{\textrm{PT}}/J$ for an $N=7$ open chain as a function of $(p,q)$, but the results remain the same for any chain size $N>2$. 
Ignoring the trivial Hermitian case ($p=q$; black, filled circles), the threshold results can be grouped into three categories: 
\begin{align}
    \label{eq:adj}
    &\textrm{Adjacent sites } (|p-q|=1):\gamma_{\textrm{PT}}= 0 \textrm{ (red circles)};\\
    \label{eq:edge}
    &\textrm{Edge sites } (|p-q|>1):\gamma_{\textrm{PT}}=J/4 \textrm{ (blue circles)};\\
    \label{eq:bulk}
    & \textrm{Bulk sites } (|p-q|>1):\gamma_{\textrm{PT}}=J/2 \textrm{ (green circles)}. 
\end{align}
When periodic boundary conditions are imposed on Eq.(\ref{eq:h0}), the "edge sites" category, Eq.(\ref{eq:edge}), disappears; the threshold is zero when the perturbations $\sigma^{\pm}$ are on adjacent sites and $\gamma_{\textrm{PT}}=J/2$ when they do not share a bond. 
These results are robust with respect to the number of spins ($N>2$), (open or periodic) boundary conditions, or the distance $|p-q|\geq 2$ and the locations of the two sites along the chain. 
This surprising nonzero threshold implies that the $\mathcal{P}$-operator is not the spatial reflection, $k\leftrightarrow N+1-k$. 
Indeed, since the Hamiltonian $H_\textrm{eff}(J,h_z|\gamma)$ is purely real, its antilinear symmetry can be chosen as $\mathcal{P}=\mathbbm{1}_N$ and $\mathcal{T}=*$ (complex conjugation). 

To understand the mechanism of $\mathcal{PT}$-symmetry breaking under exceptional perturbations, we show the flow of eigenvalues $\Re(E)$ (blue lines) and $\Im(E)$ (red lines) as a function of $\gamma/J$ in Figs.~\ref{fig:allpanels}(b)-(f). 
Since the eigenvalues occur in complex-conjugate pairs, it is sufficient to plot $\Im(E)>0$. 
When the potentials are maximally separated, $(p,q)=(1,N)$, starting from $N$ bands with varying degeneracies, a set of central bands undergo level attraction and develop imaginary parts at $\gamma=J/4$, Fig.~\ref{fig:allpanels}(b). 
The ground state (or its particle-hole symmetric counterpart) does not participate in $\mathcal{PT}$-symmetry breaking. 
Figure~\ref{fig:allpanels}(c), with $(p,q)=(6,4)$ shows that for bulk, non-adjacent sites, again, $\mathcal{PT}$-symmetry breaks with multitude of bands across the energy-level spectrum at $\gamma=J/2$. 
The trivial case of a Hermitian perturbation, $p=1=q$ shows expected linear level-splitting, Fig.~\ref{fig:allpanels}(d). 
When the perturbation sites share a bond, $(p,q)=(2,1)$ linearly increasing $\Im(E)$ signal the zero threshold, Fig~\ref{fig:allpanels}(e). 
We note that the bands developing complex eigenvalues are neither particle-hole symmetric nor at the bottom or the top. 
Lastly, when the edge perturbation sites are not maximally separated, $(p,q)=(6,1)$, the flow of eigenvalues is different, Fig.~\ref{fig:allpanels}(f), from the results in Fig.~\ref{fig:allpanels}(a). 

Next, we replace the $\Gamma^{+}_{pq}$ potential by 
\begin{align}
    \label{eq:gm}
    \Gamma^{-}_{pq}=\gamma(\sigma^{+}_p-\sigma^{-}_q)
\end{align}
which reduces to an anti-Hermitian term $\Gamma^{-}_{pp}=i\gamma\sigma_y$ when $p=q$. 
When $p\neq q$, the $\mathcal{PT}$-threshold $\gamma_{\textrm{PT}}(p,q)=\gamma_{\textrm{PT}}(q,p)$ is given by Eqs.(\ref{eq:adj})-(\ref{eq:bulk}). 
When $p=q$, the resulting threshold $\gamma_{\textrm{PT}}(p,p)=0$ for a bulk site, whereas $\gamma_{\textrm{PT}}=J/4$ for an edge site. 
Once again, these results are robust against the number of spins $N>2$, the distance $|p-q|>1$ between the perturbations, and nature of boundary conditions. 
Here, too, since Eq.(\ref{eq:gm}) has purely real entries, $\mathcal{PT}=\mathbbm{1}_N*$ gives the corresponding anti-linear symmetry.  

The simple expressions for the $\mathcal{PT}$-threshold, Eqs.(\ref{eq:adj})-(\ref{eq:bulk}), hint at an analytical solution. 
At this point, it is important to recall that the spectrum of the Hamiltonian Eq.(\ref{eq:h0}) is traditionally obtained by using the Jordan-Wigner transformation to map the problem onto non-interaction fermions~\cite{Sachdev2009,Fradkin2013}. 
Under this mapping, however, the exceptional perturbations $\sigma^+_p,\sigma^{-}_q$ create non-Hermitian, fermionic string operators, thereby rendering such an approach useless. 


\begin{figure}
    \centering
    \includegraphics[width=\columnwidth]{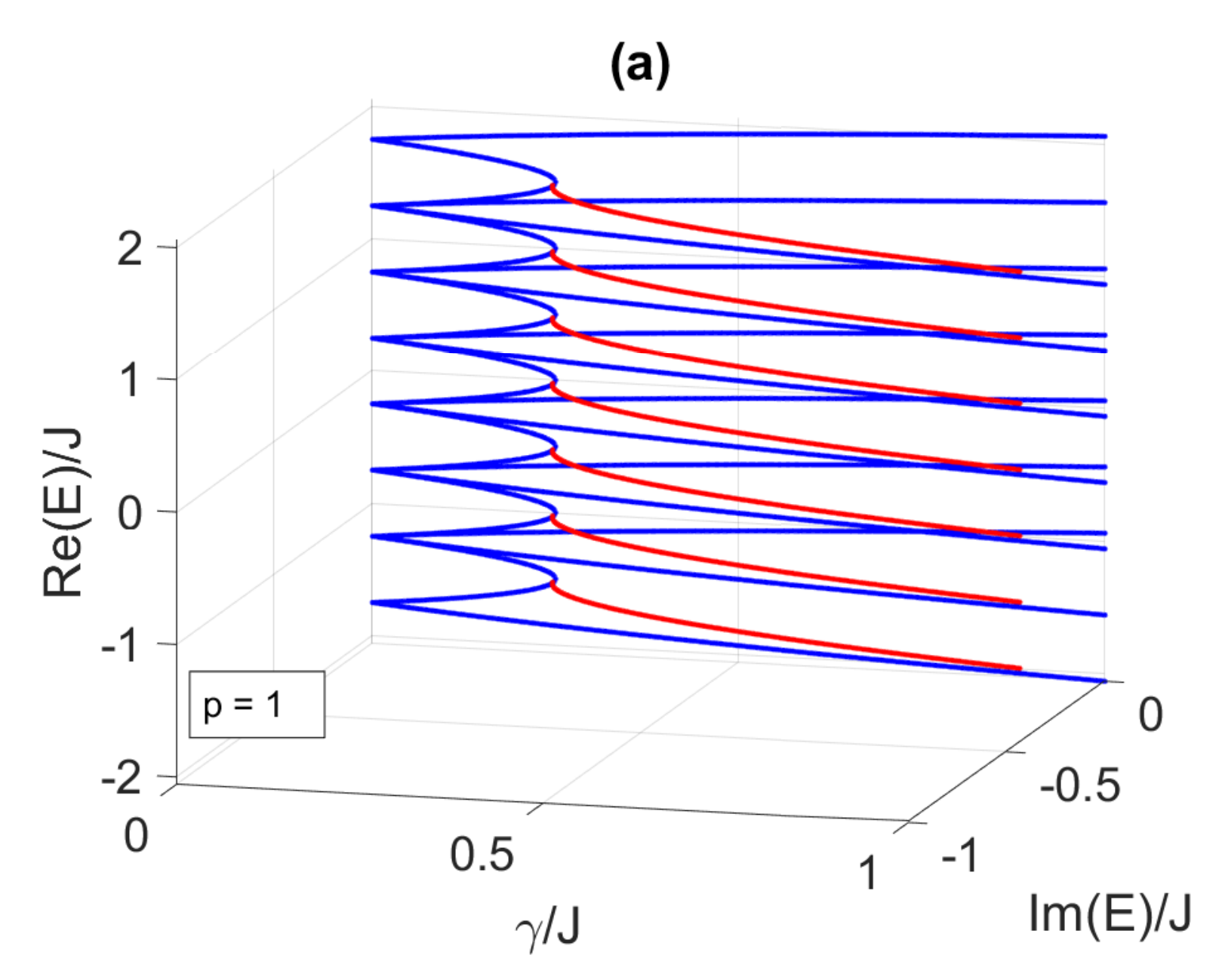}
    \includegraphics[width=\columnwidth]{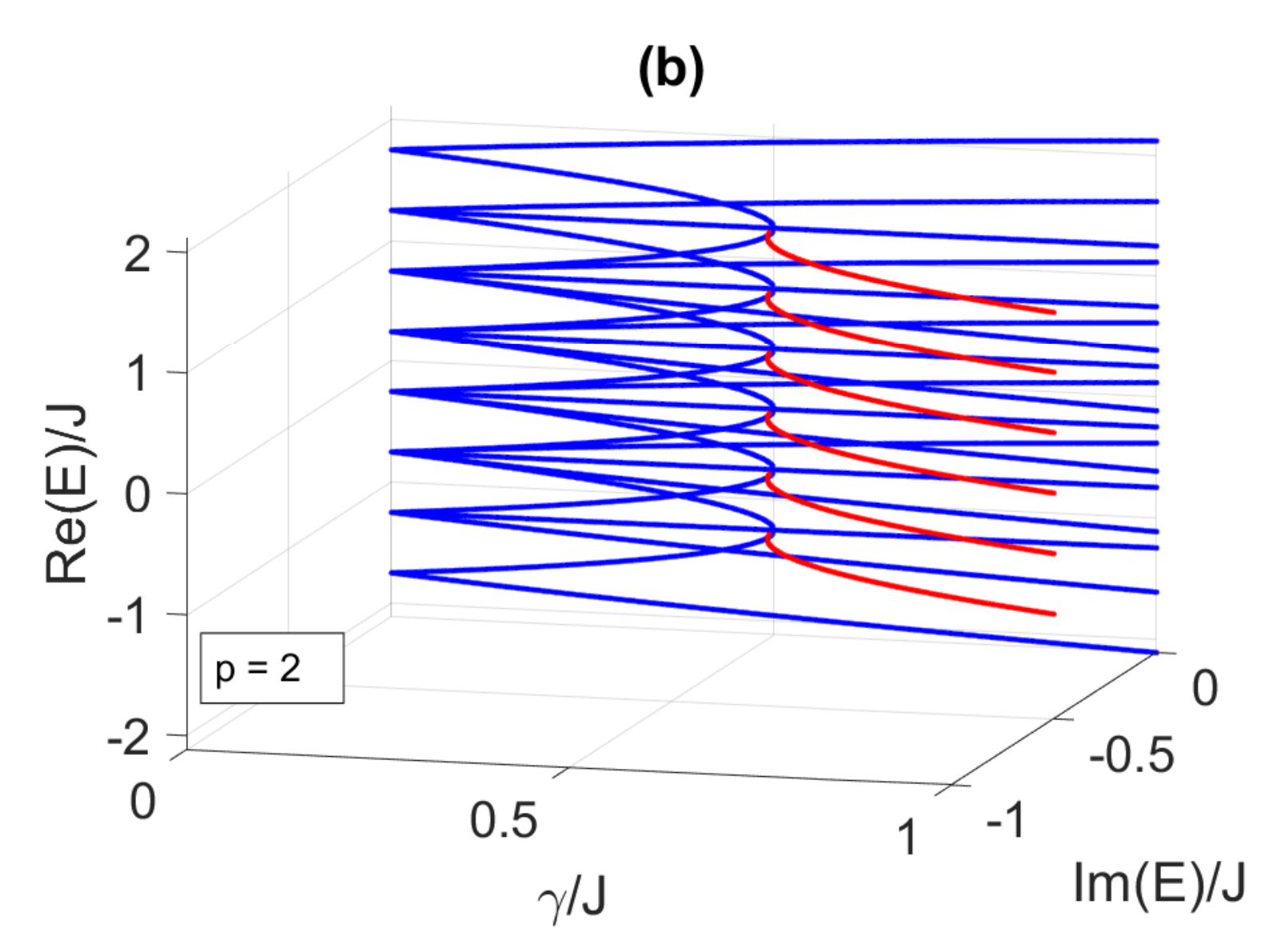}
    \caption{
        Flow of eigenvalues $E(\gamma)/J$ for an 8-spins chain with potential $\gamma\sigma^{+}$ on one site.
        Blue lines are used for $E(\gamma)/J$ that are purely real and red lines are used for $E(\gamma)/J$ that have an imaginary part
        At $\gamma=0$, the system has 8 particle-hole symmetric bands with varying degeneracies that account for the total $2^8=256$ eigenvalues. 
        (a) When the site is at the edge, $\Im(E)>0$ emerge past the threshold $\gamma_{\textrm{PT}}=J/4$. 
        (b) For a bulk site, $p=3$, the complex-conjugate eigenvalues occur past the threshold $\gamma_{\textrm{PT}}=J/2$.
    }
    \label{fig:onesite}
\end{figure}

\subsection{Single-site Perturbations \texorpdfstring{($h_{z} = 0$)}{}}
\label{subsec:one}

Inspired by the repeating structure of bands in Fig.~\ref{fig:allpanels}, and the finite $\mathcal{PT}$-threshold obtained in the anti-Hermitian limit of Eq.(\ref{eq:gm}), we now consider the Ising spin chain with a single-site perturbation,
\begin{align}
    \label{eq:gammaone} 	
    \Gamma_p(\gamma_{+},\gamma_{-})= \gamma_{+}\sigma^{+}_p+\gamma_{-}\sigma^{-}_p,
\end{align}
where $\gamma_{\pm}\in\mathbb{R}$ denote the strengths of (exceptional) non-Hermiticities $\sigma^\pm_p$ that act on the spin at site $p$. 
Starting with the case $\gamma_{-}=0$, the $\mathcal{PT}$-breaking threshold for the Hamiltonian $H_0+\Gamma_p(\gamma,0)$ is given by
\begin{align}
    \label{eq:gpone}
    \gamma_{\textrm{PT}}(p)&=\left\{\begin{array}{cc} 
        J/4 & \textrm{ Edge case, }\\
        J/2 & \textrm{ Bulk case. }
    \end{array}\right.
\end{align}

Figure \ref{fig:onesite} shows the evolution of the energy-spectra of an $N=8$ chain as a function of $\gamma$ when the sole-perturbation $\sigma^{+}$ is on the edge site (a) and the bulk site, $p=2$ (b). 
These results have many features common with the eigenvalue flows in Fig.~\ref{fig:allpanels}. 
Specifically, we see that starting with $N$ particle-hole symmetric bands at $\gamma=0$ that the $\mathcal{PT}$ breaking occurs at a threshold equal to $J/4$ or $J/2$ respectively, but the ground-state eigenvalue does not become complex. 
Since a unitary basis-change can map $\sigma_y\rightarrow -\sigma_y$ without changing interaction term in $H_0$, Eq.(\ref{eq:h0}), the threshold results for a $\Gamma_{p}(0,\gamma)$-perturbation are the same as in Eq.(\ref{eq:gpone}).

\begin{figure}
    \centering
    \vspace{-5mm}
    \includegraphics[width=\columnwidth]{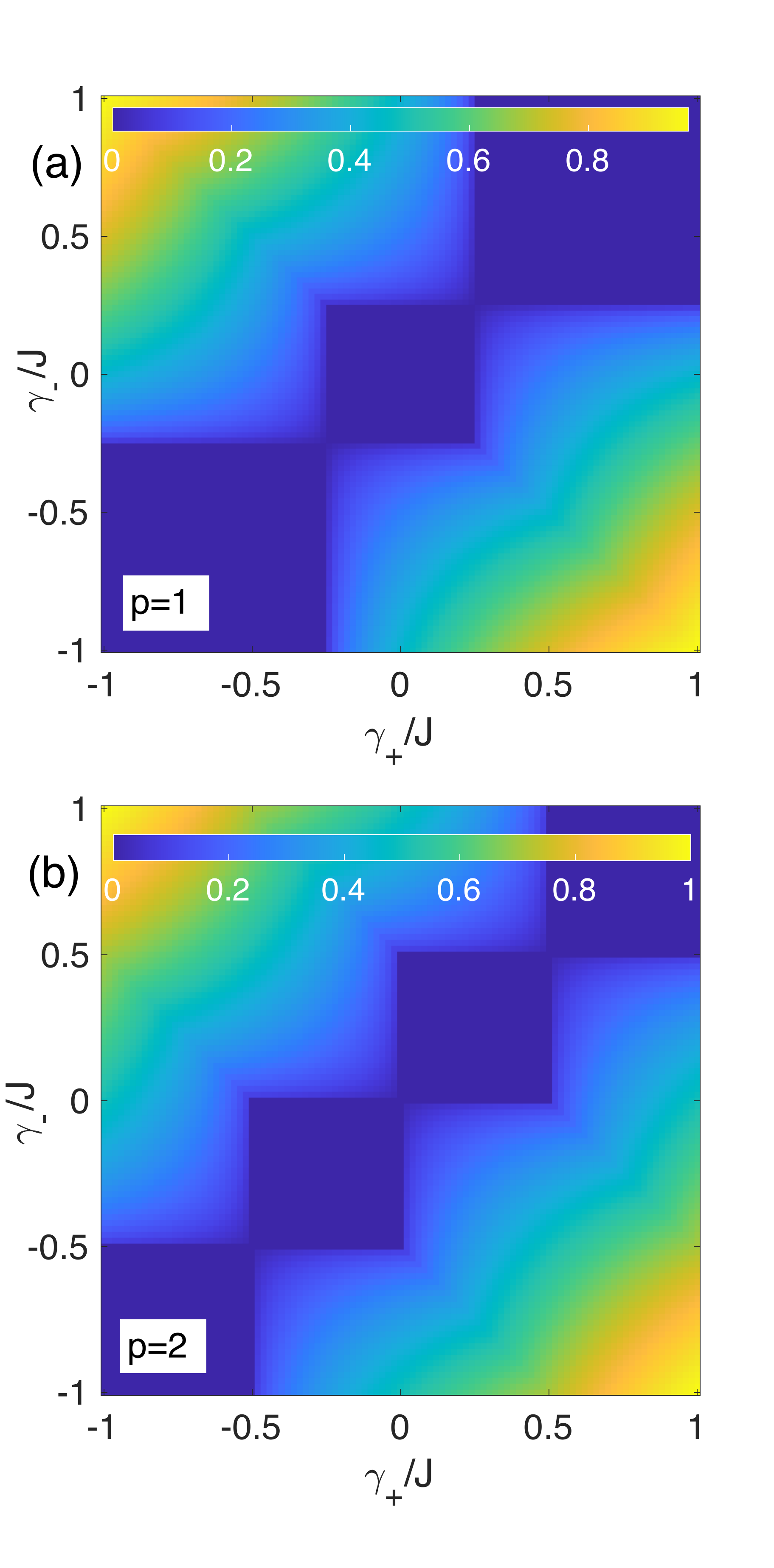}
    \vspace{-10mm}
    \caption{
        $\mathcal{PT}$-phase diagram of a 6-spins chain in $(\gamma_{+},\gamma_{-})$ plane. 
        Color denotes maximum imaginary part $\Im(E)/J$ of the eigenvalues of Hamiltonian Eq.(\ref{eq:heffplusminus}). 
        Deep blue regions indicate $\mathcal{PT}$-symmetric phase. 
        (a) When the perturbation site is at the edge ($p=1$), $\gamma_{\textrm{PT}}$ is positive along the anti-diagonal. 
        (b) When the site is in the bulk ($p=2$), the threshold is zero.
    }
    \label{fig:p12}
\end{figure}

Lastly, we consider the case where both $\gamma_\pm$ are varied. The non-Hermitian, purely real Hamiltonian is given by 
\begin{align}
    \label{eq:heffplusminus}
    H_{\textrm{eff}}=H_0+(\gamma_{+}+\gamma_{-})\sigma^x_p+i(\gamma_{+}-\gamma_{-})\sigma^y_p.
\end{align}
We characterize the $\mathcal{PT}$ phase diagram in the $(\gamma_{+},\gamma_{-})$ plane by plotting the largest imaginary part of the eigenvalues of $H_{\textrm{eff}}(\gamma_{+},\gamma_{-})$ obtained via exact diagonalization (Fig.~\ref{fig:p12}). 
It indicates whether the system is in the $\mathcal{PT}$-symmetric phase ($\max\Im(E)=0$; deep blue) or $\mathcal{PT}$-symmetry broken phase ($\max\Im(E)>0$; other colors), and quantifies the amplification rate for the $\mathcal{PT}$-broken eigenstates. 
Along the diagonal $\gamma_{+}=\gamma_{-}$, $H_{\textrm{eff}}$ is Hermitian and the spectrum is always real. 
Along the other diagonal, given by $\gamma_{+}+\gamma_{-}=0$, the perturbation (\ref{eq:gammaone}) is anti-Hermitian. 
In this case, we obtain a positive threshold for the edge case (Fig.~\ref{fig:p12}(a)), while the threshold is zero for the bulk case (Fig.~\ref{fig:p12}(b)), as seen in Sec.~\ref{subsec:two}. 

The $\mathcal{PT}$ phase-diagram in Fig.~\ref{fig:p12} is symmetric under individual reflections across the two diagonals. 
Since $\Im(E)(\gamma_{+},\gamma_{-})$ is an even function of the strength of the $i\sigma^y$ term in Eq.(\ref{eq:heffplusminus}), reflection symmetry along the main diagonal is expected. 
Reflection symmetry along the anti-diagonal, on the other hand, arises because the Hermitian term in $\Gamma_{p}$, Eq.(\ref{eq:gammaone}), commutes with the Ising interaction term $H_0$.  


\subsection{Effect of Nonzero Transverse Field on Two-site Perturbations \texorpdfstring{($h_{z} \neq 0$)}{}}
\label{subsec:nonzeroh}

When the transverse field $h_z$ is introduced, the Hamiltonian $H_{\textrm{eff}}$ contains three mutually non-commuting pieces, one for each Pauli matrix. Since the $\mathcal{PT}$-threshold results depend only on $h_z$, without loss of generality, we choose $h_z>0$. Here, we consider the fate of Hamiltonian Eq.(\ref{eq:heff}) where potentials $\sigma^{\pm}$ are introduced on sites $p,q$ respectively. Apart from the trivial Hermitian case ($p=q$), the behavior of the threshold $\gamma_{\textrm{PT}}$ can be categorized as
\begin{align}
    \label{eq:adj}
    \textrm{Adjacent sites: } &\gamma_{\textrm{PT}}(h_z)=A_1 h_z,\\
    \label{eq:bedge}
    \textrm{Both edge sites: } &\gamma_{\textrm{PT}}(h_z)=(J/4)+A_2 h_z,\\
    \label{eq:oedge}
    \textrm{One edge site: } &\gamma_{\textrm{PT}}(h_z)=(J/4)\delta_{h,0}+A_3 h_z,\\
    \label{eq:bothbulk}
    \textrm{Bulk sites: } &\gamma_{\textrm{PT}}(h_z)=(J/2)\delta_{h,0}+A_4 h_z,
\end{align}
where $A_k$ are configuration-dependent parameters. This behavior also persists when the non-Hermitian perturbation is changed to 
\begin{align}
    \label{eq:gpp}
    \Gamma^{'}_{pq}(\gamma)\equiv\gamma(\sigma^{+}_p+\sigma^{+}_q). 
\end{align}

\begin{figure}
    \centering
    \includegraphics[width=\columnwidth]{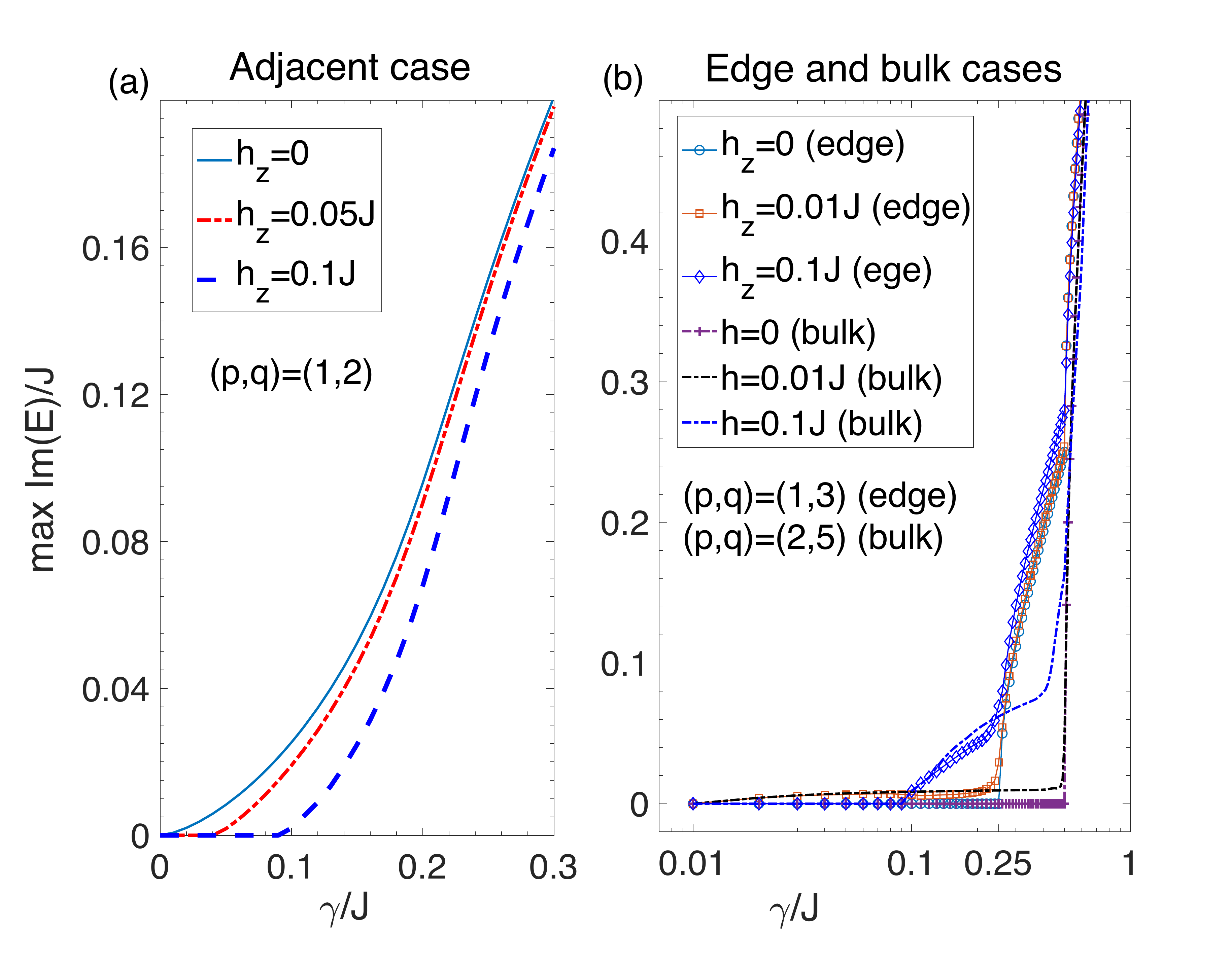}
    \caption{
        Threshold $\gamma_{\textrm{PT}}(h_z)$ for Hamiltonian Eq.(\ref{eq:heff}) with $N=9$ spins and exceptional perturbations on two sites. (a) For perturbation $\Gamma^{+}_{pq}$, adjacent sites with zero threshold develop a finite threshold $\propto |h_z|$. 
        This is signified by $\max\Im(E)=0$ regions that emerge at small $\gamma$ when $h_z\neq 0$. 
        (b) For edge sites, the $h_z=0$ threshold at $\gamma/J=0.25$, Eq.(\ref{eq:edge}), is suppressed to vanishingly small values when $h_z=0^+$ and increases with $h_z$ thereafter (solid lines with symbols). 
        For bulk sites, the threshold at $\gamma/J=0.5$, Eq.(\ref{eq:bulk}), is also suppressed to zero for $h_z=0^{+}$ and increases with $h_z$ (dot-dashed lines).
    }
    \label{fig:hz1}
\end{figure}

Figure~\ref{fig:hz1} shows the typical dependence of $\max\Im(E)(\gamma)$ on the transverse field $h_z$ for Hamiltonian $H_{\textrm{eff}}=H_0+\Gamma^{'}_{pq}$. For adjacent sites, $p=q\pm 1$, the zero threshold at $h_z=0$ is lifted to values proportional to $h_z$. This is indicated by broadening of the region where $\max\Im(E)=0$ as $\gamma$ is increased from zero (Fig.~\ref{fig:hz1}(a)). For non-adjacent cases, if one of the sites is at the edge, the $h_z=0$ threshold is given by $\gamma_{\textrm{PT}}=J/4$. It is suppressed to zero with the introduction of the transverse field. As $h_z$ increases, the threshold also increases. Similar behavior is observed for $\max\Im(E)(\gamma)$ when both sites are in the bulk (Fig.~\ref{fig:hz1}(b)).  We have verified that these results hold for spin chains up to size $N=12$.

\begin{figure}[h]
    \centering
    \includegraphics[width=\columnwidth]{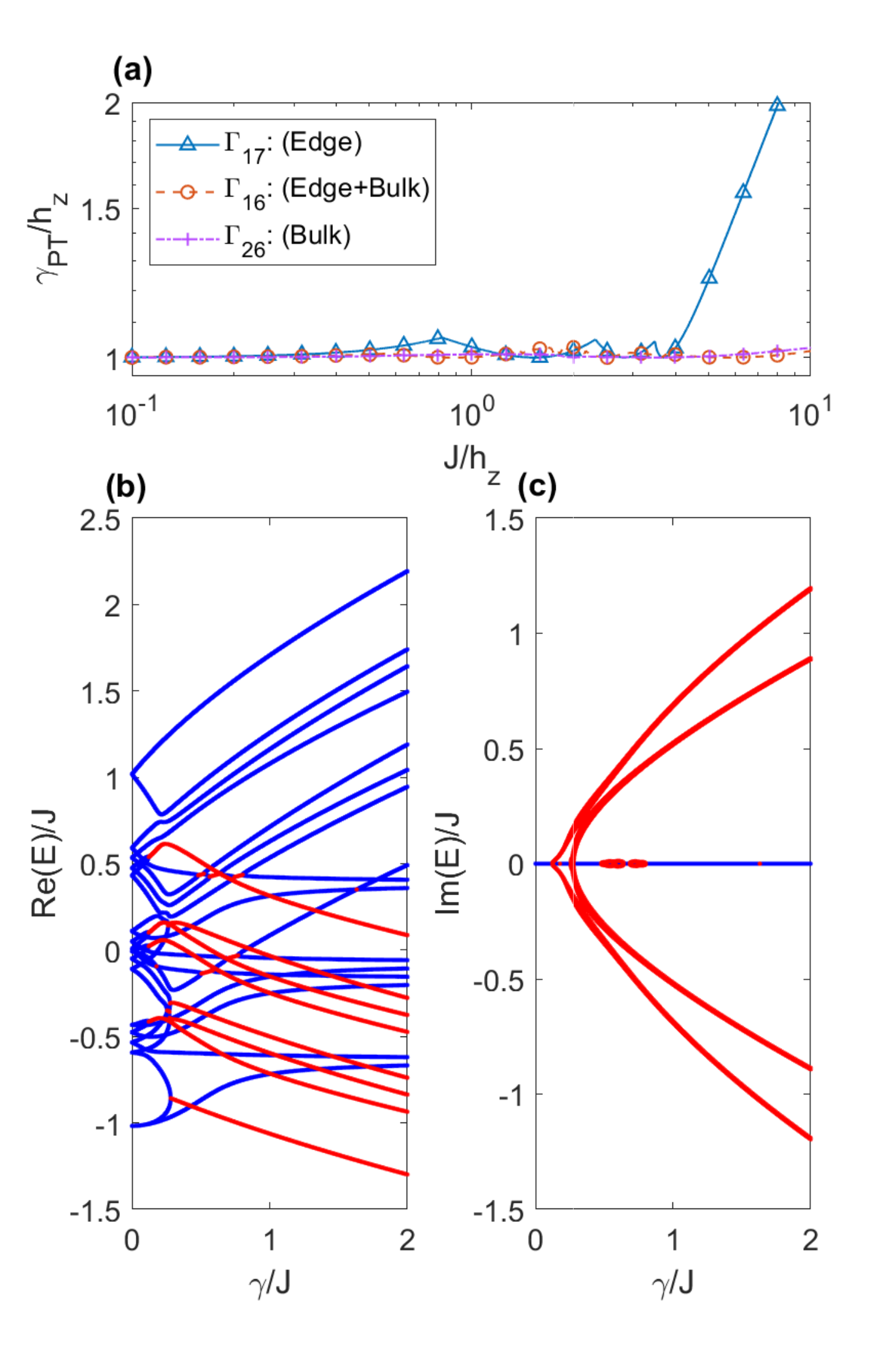}
    \caption{
        (a) Threshold $\gamma_{\textrm{PT}}(h_z)$ for Hamiltonian Eq.(\ref{eq:heff}) with $N=7$ spins, now normalized vs the transverse field $h_{z}$.
        Though $\gamma_{\textrm{PT}}(h_z)$ depends on the exact values of $p$, $q$, $J$, and $h_{z}$ the threshold is never less than $h_{z}$.
        Flow of (b) $\text{Re}(E)(\gamma/J)$ and (c) $\text{Im}(E)(\gamma/J)$ with $N=5$ spins at $h_{z}/J = 0.1$.
        The eigenvalues break in 3 distinct groups, two of which diverge but have all elements break simultaneously, and the last consists of recombinant zones that occur sporadically along $\gamma/J$.
        The group that breaks first is dependent on $p$, $q$, $J$, and $h_{z}$.
    }
    \label{fig:Jvar}
\end{figure}

Next, we consider the threshold in the weak-coupling limit, $J\ll h_z,\gamma$, where the Hamiltonian becomes
\begin{align}
    \label{eq:weak}
    H\approx-\frac{h_z}{2}\sum_{i=1}\sigma_i^{z}+\gamma(\sigma^{+}_p+\sigma^{-}_q)
\end{align}
Since the perturbations $\gamma\sigma^{\pm}=\gamma\sigma^{x}\pm i\gamma\sigma^{y}$ are already at the EP, adding a nonzero, Hermitian transverse field displaces the EP eigenvalues onto the real axis, $0=\sqrt{\gamma^2+(i\gamma)^2}\rightarrow \pm|h_z|/2$. Therefore, in the limit $J\rightarrow 0$, the $\mathcal{PT}$ threshold for Eq.\eqref{eq:weak} diverges. For intermediate values of $J/h_z$, we find that the threshold remains finite $\gamma_\mathrm{PT}\sim h_z$, albeit dependent on the configuration. Figure~\ref{fig:Jvar}(a) shows this behavior through (representative) results for an $N=7$ spin chain with configurations listed in \eqref{eq:bedge}-\eqref{eq:bothbulk}. Figure~\ref{fig:Jvar}(b) shows the eigenvalue flows for $h_z=J$ as a function of $\gamma$. The particle-hole symmetry of the spectrum in Fig.~\ref{fig:allpanels} is destroyed by a nonzero transverse field, the degeneracies are also lifted, and complex eigenvalues now occur at different values of $\gamma$ for different sets of levels.  

\subsection{Effect of Nonzero Transverse Field on Single-site Perturbation \texorpdfstring{($h_{z} \neq 0$)}{}}
\label{subsec:nonzerohsingle}

\begin{figure}[h]
    \centering
    \includegraphics[width=\columnwidth]{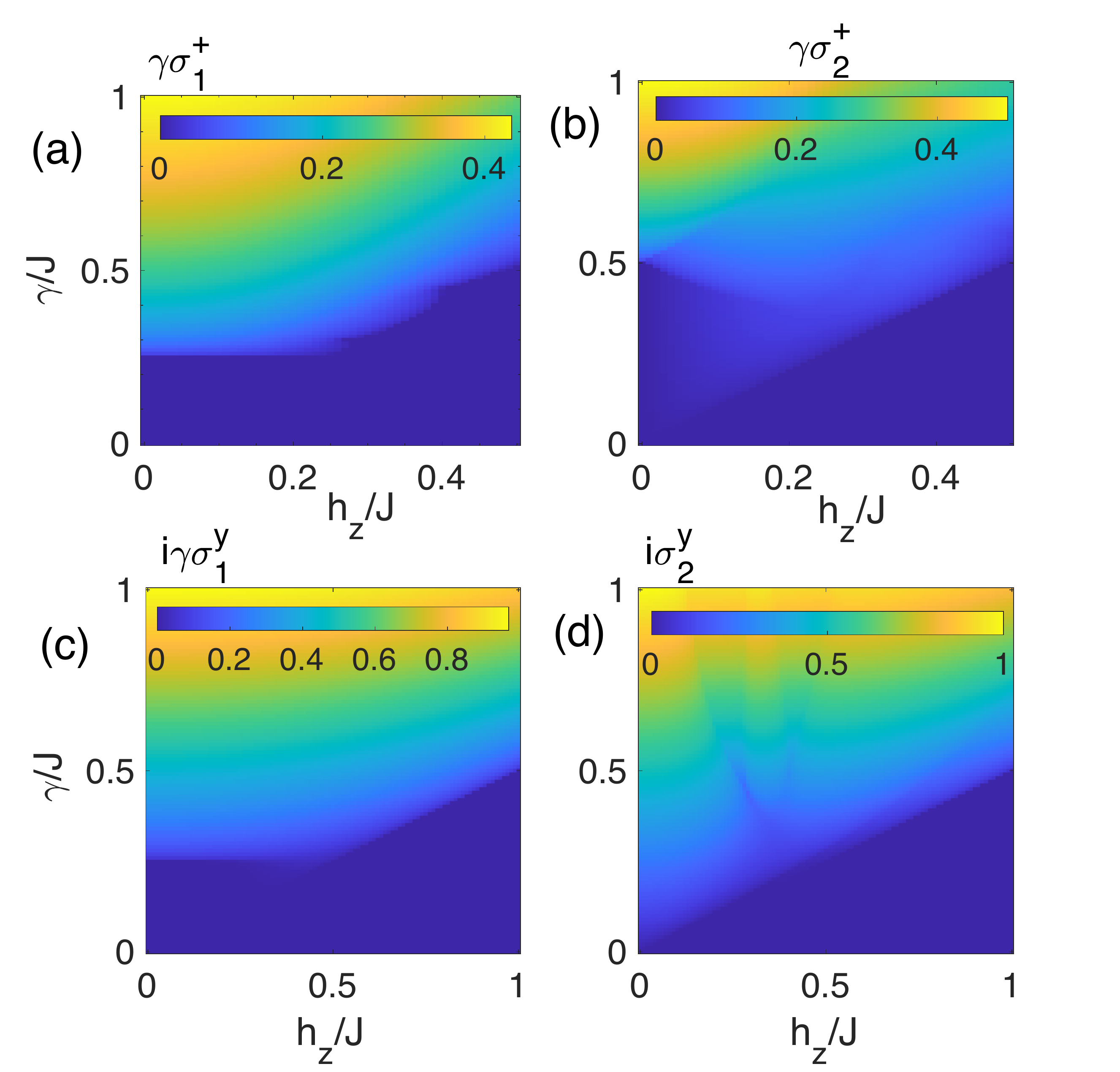}
    \caption{
        Evolution of $\max\Im(E)(\gamma,h_z)$ of the Hamiltonian Eq.(\ref{eq:heffplusminus}) with $N=7$ spins. 
        Deep blue regions ($\max\Im(E)=0$) indicate $\mathcal{PT}$-symmetric phase. 
        (a) For edge-site perturbation $\gamma\sigma^{+}_1$, the threshold increases from $J/4$ with increasing $h_z$. 
        (b) For the same perturbation in the bulk, threshold increases from $\gamma_{\textrm{PT}}(h=0^+)=0$ while its value is $J/2$ at $h_z=0$. 
        (c) same as (a) for anti-Hermitian, edge perturbation. 
        (d) same as (b) for anti-Hermitian, bulk-site perturbation, where the no-field threshold is zero.
    }
    \label{fig:hz2}
\end{figure}

Lastly, we investigate $h_z$-dependence of $\gamma_{\textrm{PT}}$ for the single-site perturbation model, Eq.(\ref{eq:heffplusminus}) with $\gamma_{-}=0$, by tracking the maximum imaginary part of its eigenvalues, $\max\Im(E)(\gamma,h_z)$. 
Fig.~\ref{fig:hz2}(a) shows that for an edge-perturbation, starting from $J/4$ the threshold continuously increases with $h_z$. 
In contrast, when the exceptional potential $\gamma\sigma^{+}$ is on an interior site, the threshold $J/2$ at $h_z=0$ is suppressed to vanishingly small values for $h_z\rightarrow 0$ before increasing linearly with $h_z$ (Fig.~\ref{fig:hz2}(b)). 
When the edge-site potential is purely anti-Hermitian, starting from $J/4$, the threshold further increases continuously with $h_z$ (Fig.~\ref{fig:hz2}(c)). 
Figure~\ref{fig:hz2}(d) shows that when the anti-Hermitian potential $i\gamma\sigma^y$ is in the bulk, the zero threshold at $h_z=0$ is linearly lifted. 
Thus, the transverse field can strengthen or weaken the $\mathcal{PT}$-symmetric phase. 



\section{Symmetries and the \texorpdfstring{$\mathcal{PT}$}{PT}-breaking threshold}
\label{sec:analytical}

The simple, $N$-independent results for the $\mathcal{PT}$-breaking threshold for a quantum Ising chain in the absence of a transverse field hint at an analytical solution. 
The robustness of that threshold $\gamma_{\textrm{PT}}$ points to the possibility of investigating the interplay between the $h_z/J$-driven quantum phase transition and the $\gamma/J$-driven $\mathcal{PT}$ symmetry breaking transition. 
Here, we discuss the analytical solution for $h_z=0$.

Consider the zero-field model with a single-site perturbation $\Gamma_p(\gamma,0)$, Eq.(\ref{eq:gammaone}). 
The eigenstates of $H_0$, Eq.(\ref{eq:h0}), can be written as $\ket{\psi}=\ket{\pm_{1}}\otimes\ket{\pm_2}\cdots\ket{\pm_N}$ where $\sigma^x_m\ket{\pm_m}=\pm\ket{\pm_m}$ are the symmetric (anti-symmetric) eigenstates at site $m$. 
For a perturbation on site $p$, we consider an eigenstate ansatz as 
\begin{align}
    \ket{\phi}\equiv \ket{\pm_1}\cdots\ket{\hat{n}_p}\cdots\ket{\pm_N},
\end{align}
where $\ket{\hat{n}_p}$ denotes the spin state at the perturbation site. 
The eigenvalue equation satisfied by the state $\ket{\phi}$ becomes
\begin{align}
    \label{eq:hp}
    H_p\ket{\hat{n}_p}&=\left[h_x\sigma^{x}+i\frac{\gamma}{2}\sigma^{y}\right]\ket{\hat{n}_p}=E_p\ket{\hat{n}_p},\\
    \label{eq:hx}
    h_x&=- \frac{J}{4}\bra{\phi}\sigma^x_{p-1}+\sigma^x_{p+1}\ket{\phi}+\frac{\gamma}{2},
\end{align}
where one of the $p\pm 1$ terms is absent when the location $p$ is at the edge. 
The $2\times 2$ Hamiltonian $H_p$ Eq.(\ref{eq:hp}) undergoes $\mathcal{PT}$-symmetry breaking when the strength of the imaginary field is equal to that of the real field, i.e $h_x=\pm\gamma/2$. 
This gives Eq.(\ref{eq:gpone}) as the threshold result. 
A similar analysis can be carried out for other exceptional potentials, including two-site potentials, Eq.(\ref{eq:G}), when the two sites are not adjacent. When the two sites are adjacent, a similar reduction to a $4\times 4$ Hamiltonian gives the zero threshold, Eq.(\ref{eq:adj}). The interaction contribution to the effective field in Eq.\eqref{eq:hx} vanishes in states where the neighboring spins $p\pm 1$ have opposite projections. For such states, effective Hamiltonian $H_p$ remains at the exceptional point, producing $\gamma$-independent flat bands in the energy spectrum that never become complex. This unusual behavior results solely from our choice of exceptional non-Hermiticities $\gamma\sigma^{\pm}$ that generate no energy scale.

It is a useful exercise to think about how the threshold analysis presented here would look like in the Jordan-Wigner fermions language. When $h_z=0$, the single or two-site non-Hermiticities will lead to $p,q$-dependent fermionic strings, all of which lead to simple threshold answers. One might imagine obtaining different energy shifts through a perturbative analysis. However, it is known that the perturbation theory cannot be used to predict the $\mathcal{PT}$-threshold as its radius of convergence is exactly at the boundary between the $\mathcal{PT}$-symmetric and broken regions~\cite{Moiseyev1980,Klaiman2008}. Adding a transverse field $h_z$ only changes the fermions that diagonalize the Hermitian Hamiltonian through a Bogoliubov transform. Thus, the relative insensitivity of the threshold $\gamma_\mathrm{PT}$ to configuration details is a common feature of system with or without the transverse field.

Note that although the two-site perturbation was motivated by splitting a Hermitian term into two Jordan-normal-form terms, symmetries in the $h_z=0$ case map $\sigma^{+}\leftrightarrow\sigma^{-}$ under a local, unitary transformation on the site of the potential. 
This equivalence between the two potentials is another reminder that in systems with bounded eigenvalue spectrum, "gain" and "loss" are not equivalent to raising and lowering operators. 
Additional unitary-equivalent terms such as $\gamma\sigma^{+}\leftrightarrow-\gamma\sigma^{+}$ or $h_z\leftrightarrow -h_z$ were already taken into account when obtaining the $\mathcal{PT}$ threshold results. 


\section{Discussion}
\label{sec:disc}

In this paper, we have developed a new class of $\mathcal{PT}$-symmetric quantum Ising models with $N$ spins, where the non-Hermitian potentials are confined to one or two sites and the $\mathcal{PT}$-breaking threshold is independent of $N>2$. In most traditional models, the non-Hermiticity is introduced by changing model parameters from real to complex. That means the number of sites with non-Hermiticity is proportional to $N$, something that is virtually impossible to implement in coupled-qubits realizations of a quantum spin chain. Therefore, with experimental feasibility in mind, we have chosen localized non-Hermiticities. Our second deliberate choice is that instead of commonly used anti-Hermitian potentials (obtained by changing a parameter from real to purely imaginary), we have used perturbations that, by themselves, do not generate an energy scale and therefore keep the system at an EP in the limit $\gamma\gg J,h_z$.

Our models show that introducing a single non-Hermitian qubit in a Hermitian, quantum Ising chain gives rise to $\gamma_{\mathrm{PT}}$ that can be varied with the transverse field. With full control required over only the non-Hermitian qubit, our models provide a pathway to investigate the interplay between interaction and non-Hermitian properties. Our results remain qualitatively unchanged when the Hermitian Hamiltonian is changed from a quantum Ising model to its integer-spin counterpart or Heisenberg model with or without anisotropies. The spin-1 case, for example, is made richer by the possibility of different exceptional perturbations such as $S^{+}=(S^x+i S^y)/2$ and $S^{+2}\neq 0$. An exact diagonalization analysis is required to obtain the general threshold $\gamma_{PT}(J_{xx},J_{yy},J_{zz};{\bf h})$ as a function of the multiple, possible non-Hermiticities, and its exhaustive characterization is an open problem. 

\begin{acknowledgments}
This work is supported by ONR Grant No. N00014-21-1-2630. We thank P. Durganandini and Kater Murch for discussions.  
\end{acknowledgments}



\bibliography{referencesJM.bib}

\end{document}